\newif\ifarxiv\arxivtrue
\ifarxiv\pdfoutput=1\fi
{\def\usepackage{ws-procs9x6}}
\documentclass[12pt,a4paper]{article}

\ifarxiv\ifnum\pdfoutput=1\else
\PassOptionsToPackage{hypertex}{hyperref}
\PassOptionsToPackage{draft}{graphicx}
\usepackage{showkeys}
\fi\fi

\usepackage{amsmath,amssymb}
\usepackage[bookmarks=true,hyperfigures=true]{hyperref}
\usepackage{graphicx}
\usepackage[nosort]{cite}
\usepackage[bulletsep]{collref}

\usepackage[a4paper,text={160mm,247mm},centering]{geometry}

\def\showkeysrefformat#1{\usefont\encodingdefault\ttdefault\mddefault\updefault\tiny#1}
\makeatletter
\def\SK@@ref#1>#2\SK@{%
 {\@inlabelfalse\leavevmode\vbox to\z@{%
 \vss\SK@refcolor\rlap{\vrule\raise .75em%
  \hbox{\showkeysrefformat{#2}}}}}}
\makeatother

\allowdisplaybreaks[3]

\numberwithin{equation}{section}

\usepackage[font=small,labelfont=bf,width=0.85\textwidth]{caption}

\expandafter\def\expandafter\bfseries\expandafter{\bfseries\ifmmode\else\boldmath\fi}
\expandafter\def\expandafter\mdseries\expandafter{\mdseries\ifmmode\else\unboldmath\fi}
\expandafter\def\expandafter\normalfont\expandafter{\normalfont\ifmmode\else\unboldmath\fi}

\RequirePackage{verbatim}

\makeatletter
\newwrite\bibinl@out
\newenvironment{bibtex}[1][\jobname]{%
  \immediate\openout\bibinl@out #1.bib
  \immediate\write\bibinl@out{\@percentchar generated from `\jobname' starting line \the\inputlineno^^J}%
  \def\verbatim@processline{\immediate\write\bibinl@out{\the\verbatim@line}}%
  \@bsphack\let\do\@makeother\dospecials\catcode`\^^M\active\verbatim@start
}%
{\immediate\closeout\bibinl@out\@esphack}
\makeatother

\usepackage{fixmath}

\ifx\genfrac\sdflkaj\else\fi
\newcommand{\sfrac}[2]{{\textstyle\frac{#1}{#2}}}
\newcommand{\half}{\sfrac{1}{2}}

\newcommand{\quarter}{\sfrac{1}{4}}

\newcommand{\Complex}{\mathbb{C}}
\newcommand{\Integer}{\mathbb{Z}}

\newcommand{\alg}[1]{\mathfrak{#1}}
\newcommand{\grp}[1]{\mathrm{#1}}

\newcommand{\matr}[2]{\left(\begin{array}{#1}#2\end{array}\right)}

\newcommand{\lrbrk}[1]{\left(#1\right)}
\newcommand{\bigbrk}[1]{\bigl(#1\bigr)}

\newcommand{\comm}[2]{[#1,#2]}

\DeclareMathOperator{\tr}{tr}
\DeclareMathOperator{\str}{str}

\newcommand{\order}[1]{\mathcal{O}(#1)}
\newcommand{\gen}[1]{\mathrm{#1}}
\newcommand{\dual}{\mathord{\ast}}
\newcommand{\map}{\Sigma}
\newcommand{\cmap}{\Pi}
\newcommand{\trans}{{\scriptscriptstyle\mathsf{T}}}

\newcommand{\nln}{\nonumber\\}

\def\[{\begin{equation}}
\def\]{\end{equation}}

\providecommand{\href}[2]{#2}

\makeatletter
\def\mr@ignsp#1 {\ifx\:#1\@empty\else #1\expandafter\mr@ignsp\fi}%
\newcommand{\multiref}[1]{\begingroup
\xdef\mr@no@sparg{\expandafter\mr@ignsp#1 \: }%
\def\mr@comma{}%
\@for\mr@refs:=\mr@no@sparg\do{\mr@comma\def\mr@comma{,}\ref{\mr@refs}}%
\endgroup}
\renewcommand{\eqref}[1]{(\multiref{#1})}
\makeatother

\makeatletter
\newcommand{\namedref}[2]{\hyperref[#2]{#1~\ref*{#2}}}
\newcommand{\secref}{\@ifstar{\namedref{Section}}{\namedref{Sec.}}}
\newcommand{\appref}{\@ifstar{\namedref{Appendix}}{\namedref{App.}}}
\newcommand{\tabref}{\@ifstar{\namedref{Table}}{\namedref{Tab.}}}
\newcommand{\figref}{\@ifstar{\namedref{Figure}}{\namedref{Fig.}}}
\makeatother

\providecommand{\hypersetup}[1]{}
\providecommand{\texorpdfstring}[2]{#1}

\hypersetup{plainpages=false}
\hypersetup{pdfpagemode=UseNone}
\hypersetup{bookmarksnumbered=true}
\hypersetup{pdfstartview=FitH}
\hypersetup{colorlinks=false}
\hypersetup{citebordercolor={.5 1 .5}}
\hypersetup{urlbordercolor={.5 1 1}}
\hypersetup{linkbordercolor={1 .7 .7}}

\makeatletter
\let\@keywords\@empty
\let\@subject\@empty
\providecommand{\keywords}[1]{\gdef\@keywords{#1}}
\providecommand{\subject}[1]{\gdef\@subject{#1}}
\def\thetitle{\@title}
\def\theauthor{\@author}
\def\thesubject{\@subject}
\def\thedate{\@date}
\def\thekeywords{\@keywords}
\makeatother
\AtBeginDocument{
\hypersetup{pdftitle={\thetitle}}%
\hypersetup{pdfauthor={\theauthor}}%
\hypersetup{pdfsubject={\thesubject}}%
\hypersetup{pdfkeywords={\thekeywords}}%
}

\title{Construction of Lax Connections\texorpdfstring{\\}{ }by Exponentiation}
\author{Niklas Beisert and Florian L\"ucker}

\begin{document}
\pdfbookmark[1]{Title Page}{title}
\thispagestyle{empty}


\vspace*{2cm}
\begin{center}%
\begingroup\Large\bfseries\thetitle\par\endgroup
\vspace{1cm}

\begingroup\scshape\theauthor\par\endgroup
\vspace{5mm}%

\begingroup\itshape
Institut f\"ur Theoretische Physik,\\
Eidgen\"ossische Technische Hochschule Z\"urich\\
Wolfgang-Pauli-Strasse 27, 8093 Z\"urich, Switzerland
\par\endgroup
\vspace{5mm}

\begingroup\ttfamily
nbeisert@ethz.ch,\\
fluecker@student.ethz.ch
\par\endgroup

\vfill

\textbf{Abstract}\vspace{5mm}

\begin{minipage}{12.7cm}
We propose a method for constructing the Lax connection 
of two-dimensional relativistic integrable sigma models on coset spaces 
by means of exponentiation of a suitable operator.
We derive a simple quadratic relation that this operator
must satisfy for an entire one-parameter family of connections to be flat.
\end{minipage}

\vspace*{4cm}

\end{center}

\newpage

\section{Introduction}

Symmetries in physical models lead to conservation laws 
which make the investigation of the model somewhat more tractable. 
Integrable models have a maximum amount of conservation laws 
which can be viewed to arise from a large or even infinite number of hidden symmetries. 
The dynamics of these models is largely or completely determined by the symmetries.

The integrable structure of many integrable models of mechanics can be formulated
in terms of a Lax pair \cite{Lax:1968fm}, 
a pair of matrices $L(t),M(t)$
depending on the phase space of the mechanical system, 
such that the equations of motion can be written as
\[
\frac{dL}{dt}=\comm{M}{L}.
\]
The equation guarantees that the eigenvalues of $L(t)$ are conserved quantities. 
For sufficiently big matrices, it implies integrability of a model
with finitely many degrees of freedom.

For two-dimensional models a conventional Lax pair is not sufficient:
Firstly, one has to accommodate for the additional spatial coordinate $x$. 
Secondly, an integrable field theory model
requires infinitely many conservation laws.
The Lax pair formalism can be extended 
to a Lax connection $A(\lambda)$ 
on the two-dimensional spacetime. 
It must obey the flatness condition
\[
dA(\lambda)+A(\lambda)\wedge A(\lambda)=0
\]
for all values of the spectral parameter $\lambda$.
It can be connected to the above Lax pair by identifying 
$L$ with the monodromy matrix
and $M$ with the time-component of $A$
\footnote{Strictly speaking this identification requires
periodic boundary conditions for $x$.}
\[
L(t,\lambda)=\mathrm{P}\exp\int dx\,A_x,
\qquad
M(t,\lambda)=A_t(x_0,t,\lambda).
\]
Now we have a Lax pair for every value of the spectral parameter $\lambda$,
and effectively there are infinitely many conserved quantities. 
One may also view $\lambda$ as a substitute for the spacial 
dependence of the fields in the Lax pair. 
Ultimately, the Lax connection can be used in a variety of 
ways to construct and investigate solutions of the equations of motion.

Integrability is a hidden symmetry enhancement, and as such, it is not
straight-forward to detect. Given some two-dimensional field theory model,
how to determine integrability? 
A standard method is to make a sufficiently general ansatz for the Lax 
connection in terms of the fields, and then determine the
coefficients such that the flatness condition is satisfied. 
What makes life difficult is that the flatness condition is non-linear.
Furthermore, a single non-trivial solution will not suffice, 
integrability requires a one-parameter family of flat connections $A(\lambda)$. 
The existence of such a solution may appear like magic, 
but after all this magic is integrability. 
With some experience and inspiration one can usually 
find the Lax connection for a given model,
but this is more or less trial and error.

In this paper we propose a very direct construction 
of the Lax connection in terms of an operator $\map$
\[
A(\lambda) = \exp(\lambda \map) J.
\]
Here, $J$ is the Maurer--Cartan form of a particular
class of two-dimensional relativistic sigma models
on a group manifold or cosets thereof. 
The operator turns out to be immediately connected to the equations
of motion and to the action. 
We will also derive a necessary and sufficient condition for integrability. 
Altogether this condition can be used as a direct integrability test.
It certainly does not apply to all two-dimensional
field theory models, not even to all such sigma models.
However, a positive result of the test not only proves integrability, 
but also provides the Lax connection at the same time.
We will outline our construction, investigate the 
properties it requires,
and apply the method to a couple of interesting sigma models.

\medskip

The paper is organised as follows:
We start in \secref{sec:sigma} by introducing
two-dimensional sigma models
formulated in terms of Maurer--Cartan forms.
For a couple of well-known integrable models, 
we state the action, equations of motion and the Lax connection. 
In the following \secref{sec:map} we sketch 
our novel construction of the Lax connection by exponentiation 
in these models.
We then investigate our proposal for quite a general class 
of sigma models in \secref{sec:derivation}. 
In particular, we derive a condition for the underlying operator
of the construction which is necessary and sufficient for integrability.
We return to our sample models and a few other relevant 
integrable sigma models in \secref{sec:application}.
Finally in \secref{sec:shift}, we discuss 
an alternative interpretation of our construction as a shift operator.
We conclude in \secref{sec:concl} and give an outlook.

\section{Sigma Models on Coset Spaces}
\label{sec:sigma}

In this paper we will mainly be concerned with 
two-dimensional relativistic non-linear sigma models with 
a group manifold $\grp{G}$ or 
a coset space $\grp{G}/\grp{H}$ of Lie groups $\grp{G},\grp{H}$
as target space.
Let us briefly review their formulations in terms of
Maurer--Cartan forms, integrability in terms of
a Lax connection as well as a few examples.

\paragraph{General Framework.}

The sigma model is based on the group-valued field $g(x)\in \grp{G}$.
For coset models the field should be viewed 
modulo the equivalence relation $g\simeq gh$ with $h(x)\in \grp{H}$.
We will work with the associated Maurer--Cartan one-form $J=g^{-1}dg$
taking values in the Lie algebra $\alg{g}$ corresponding to $\grp{G}$.
The equivalence relation transforms $J$ as 
\[\label{eq:MCsim}
J\simeq J'=h^{-1}(d+J)h.
\]
Bearing in mind that $J$ obeys the Maurer--Cartan equation 
\[\label{eq:MC}
dJ+J\wedge J=0,
\] 
the form field $J$ offers an equivalent description
of the model 
up to global transformations $g(x)\mapsto g_0 g(x)$
with $g_0\in \grp{G}$
and up to potential topological issues.
Furthermore, there will be an action and 
a derived set of equations of motion which determine
the dynamics of the sigma model. 
The latter is considered integrable if there exists 
a non-trivial one-parameter family of $\alg{g}$-connections
$A(\lambda)$, $\lambda\in\Complex$,
which obeys the on-shell flatness condition 
\[\label{eq:flatness}
dA(\lambda)+A(\lambda)\wedge A(\lambda) = 0\quad
\mbox{for all }\lambda\in\Complex.
\]
Alternatively, the flatness condition can
be viewed to encode the above Maurer--Cartan equations 
(conventionally at $\lambda=0$) 
and the equations of motion 
(conventionally at $\order{\lambda^1}$).

Many relevant cosets are specified 
by an automorphism $\Integer_N$ of $\grp{G}$ 
whose fix points define the subgroup $\grp{H}$.
In these cases it is convenient to 
split up $J$ according to the $\Integer_N$-grading 
\[J=\sum_{k=0}^{N-1} J_{(k)}.\]
Consequently, the Maurer--Cartan equations split up according to the $\Integer_N$-grading
\[\label{eq:ZNMC}
dJ_{(k)}+\sum_{j=0}^{N-1} J_{(j)}\wedge J_{(k-j)}=0
\]
for all $k=0,\ldots,N-1$.
Under the equivalence relation \eqref{eq:MCsim} $J_{(0)}$ transforms 
as a $\alg{h}$-connection whereas the other $J_{(k)}$ transform 
by conjugation by $h$. Hence, physical quantities 
such as the equations of motion and the Lagrangian 
are constructed from the $J_{(k\neq 0)}$ 
and the covariant derivative $d+J_{(0)}$.

\paragraph{Symmetric Space Sigma Model.}

The simplest case is a symmetric space, i.e.\ a coset with $\Integer_2$-grading.
Here the action must be proportional to $\int \tr J_{(1)}\wedge \dual J_{(1)}$.
The resulting equations of motion along with the Maurer--Cartan equations
\eqref{eq:ZNMC} read
\begin{align}\label{eq:Z2eqs}
0&=dJ_{(0)} + J_{(0)}\wedge J_{(0)} + J_{(1)}\wedge J_{(1)},
\nln
0&=dJ_{(1)} + J_{(0)}\wedge J_{(1)} + J_{(1)}\wedge J_{(0)},
\nln
0&=d\dual J_{(1)} + J_{(0)}\wedge \dual J_{(1)} + \dual J_{(1)}\wedge J_{(0)}.
\end{align}
Coset models on symmetric spaces are integrable 
\cite{Pohlmeyer:1975nb,Luscher:1977rq,Eichenherr:1978qa,Eichenherr:1979ci}.
Integrability can be expressed by means of the Lax connection 
\[\label{eq:LaxZ2}
A(\lambda) = J_{(0)} + \half (e^{-\lambda}+e^\lambda) J_{(1)} + \half(e^{-\lambda}-e^\lambda) \dual J_{(1)},
\]
whose flatness condition \eqref{eq:flatness}
summarises all of the above equations \eqref{eq:Z2eqs}.
Note that the Lax connection is uniquely specified 
only up to reparametrisations of $\lambda\to f(\lambda)$.
Two commonly used alternative parametrisations 
are defined by $e^\lambda=z=(x+1)/(x-1)$.

\paragraph{$\Integer_3$-Coset Model.}

A less studied yet interesting case is a $\Integer_3$-coset.
Now there are two invariant terms for the action
$\int \tr J_{(1)}\wedge \dual J_{(2)}$ and $\int \tr J_{(1)}\wedge J_{(2)}$. 
For a plain sigma model any combination of the two is fine, 
but for an integrable one they must come with a relative coefficient
of $3$ for the first term \cite{Young:2005jv}.%
\footnote{An opposite relative sign leads to an equivalent integrable model.
For definiteness we will restrict our analysis to one choice.}
The corresponding equations of motion
read
\begin{align}\label{eq:Z3eqs}
0&=d\dual J_{(1)} + J_{(0)}\wedge \dual J_{(1)} + \dual J_{(1)}\wedge J_{(0)} + J_{(2)}\wedge J_{(2)},
\nln
0&=d\dual J_{(2)} + J_{(0)}\wedge \dual J_{(2)} + \dual J_{(2)}\wedge J_{(0)} - J_{(1)}\wedge J_{(1)}.
\end{align}
%
Together with the Maurer--Cartan equations
\eqref{eq:ZNMC} they are summarised 
in the flatness \eqref{eq:flatness} of the Lax connection \cite{Young:2005jv}
\begin{align}
\label{eq:LaxZ3}
A(\lambda) =\mathord{}& J_{(0)} 
+ \half (e^{-\lambda}+e^{2\lambda}) J_{(1)} 
+ \half (e^{-\lambda}-e^{2\lambda}) \dual J_{(1)} 
\nln\mathord{}&\phantom{J{(0)}}
+ \half (e^{-2\lambda}+e^{\lambda}) J_{(2)} 
+ \half (e^{-2\lambda}-e^{\lambda}) \dual J_{(2)}.
\end{align}
%
In fact, there is a generalisation to $\Integer_N$-cosets
which we shall discuss further below.

\paragraph{$\Integer_4$ Symmetric Superspace Model.}

The generalisation of symmetric spaces
to supermanifolds extends the $\Integer_2$-grading
to a $\Integer_4$-grading \cite{Berkovits:1999zq}. 
Such models,
particularly IIB superstrings on $AdS_5\times S^5$ \cite{Metsaev:1998it},
are central to the study of integrable structures within the AdS/CFT correspondence,
see \cite{Beisert:2010jr,Magro:2010jx} for reviews. 
Here the Maurer--Cartan form splits into components $J_{(0)},J_{(2)}$
of bosonic statistics and components $J_{(1)},J_{(3)}$
of fermionic statistics. 
The $\Integer_2$ statistic group is thus a subgroup 
of the $\Integer_4$ automorphism group.
There are several conceivable terms for the action, 
but in these models the action is composed
from $\int \str J_{(2)}\wedge \dual J_{(2)}$ 
and $\int \str J_{(1)}\wedge J_{(3)}$.
Integrability as well as other desirable properties
imply a relative coefficient of $-1$ between the terms 
\cite{Bena:2003wd}.
The resulting equations of motion read%
\footnote{We should point out that here and elsewhere the $\dual$-dualisation 
involves the metric of 2D spacetime. 
For models with a non-constant metric
the derivative acts non-trivially on $\dual$ itself.
Gladly, the equations of motion are typically of the form 
$d\ast J=\ldots$ with this effect implied.
Moreover, in gravity or string theory models,
a dynamical metric gives rise to additional constraints not contained 
in the usual sigma model equations of motion. 
These do not affect integrability.}
\begin{align}\label{eq:Z4eqs}
0 &=\phantom{d\dual J_{(2)}+\mathord{}}J_{(3)}\wedge \dual J_{(2)}+\dual J_{(2)}\wedge J_{(3)}-J_{(3)}\wedge J_{(2)}-J_{(2)}\wedge J_{(3)},
\nln
0&=d\dual J_{(2)} + J_{(0)}\wedge \dual J_{(2)}+\dual J_{(2)}\wedge J_{(0)} +J_{(1)}\wedge  J_{(1)} -J_{(3)}\wedge  J_{(3)},
\nln
0 &=\phantom{d\dual J_{(2)}+\mathord{}}J_{(1)}\wedge \dual J_{(2)}+\dual J_{(2)}\wedge J_{(1)}+J_{(1)}\wedge J_{(2)}+J_{(2)}\wedge J_{(1)}.
\end{align}
Again they are implied by the flatness of the following Lax connection
\[\label{eq:LaxZ4}
A(\lambda) = J_{(0)} 
+ e^{-\lambda} J_{(1)} 
+ \half(e^{-2\lambda}+e^{2\lambda}) J_{(2)}
+ \half (e^{-2\lambda}-e^{2\lambda}) \dual J_{(2)} 
+ e^\lambda J_{(3)}.
\]

\section{Exponential Map}
\label{sec:map}

The standard method to construct a Lax connection $A$ 
for models with some decomposition of the Maurer--Cartan form 
is to make a general ansatz for $A$ as a linear combination of
the $J_{(k)}$ and their duals $\dual J_{(k)}$. 
The flatness condition $dA+A\wedge A=0$ has the trivial solution $A=J$
due to the Maurer--Cartan equation \eqref{eq:MC}.
When the model is integrable, this solution
should extend to a one-parameter family $A(\lambda)$. 
As mentioned above the overall parametrisation is arbitrary.

We have chosen a parametrisation where the coefficients in $A(\lambda)$
are Laurent polynomials in $e^\lambda$. 
The fact that this is possible is somewhat curious. 
It points to the possibility that the construction of the Lax connection
involves some sort of exponential map.
More concretely, we observe that the above Lax connections can be 
constructed as
\[\label{eq:expmap}
A(\lambda)=\exp(\lambda \map)\circ J,
\]
where $\map$ is a linear operator on the space
$\alg{g}\oplus\alg{g}$ spanned by 
the one-form fields $J$ and their duals $\dual J$
viewed as elements of the Lie algebra $\alg{g}$.

For sigma models on symmetric spaces
the operator $\map$ is defined as 
\[\label{eq:MapZ2}
\map (J_{(0)})=0,\qquad
\map (J_{(1)})=-\dual J_{(1)},\qquad
\map (\dual J_{(1)})=-J_{(1)}.
\]
The action on $\dual J_{(0)}$ is not needed,
one might define $\map (\dual J_{(0)})=0$. 
Let us compute $A(\lambda)$ from the formula \eqref{eq:expmap}
where we Taylor expand the exponential.
The action of $\map^n$ reads
\[
\map^0(J)
=
J_{(0)}+J_{(1)},
\qquad
\map^{2n}(J)=J_{(1)},
\qquad
\map^{2n+1}(J)=-\dual J_{(1)}.
\]
This obviously yields
$
A(\lambda)=J_{(0)}+\cosh (\lambda)\, J_{(1)}-\sinh(\lambda)\, \dual J_{(1)}
$,
which matches precisely with \eqref{eq:LaxZ2}.

For the $\Integer_3$-coset we define the operator
\begin{align}
\map(J_{(0)})&=0,&
\map(J_{(1)})&=\sfrac{3}{2}\dual J_{(1)}+\sfrac{1}{2}J_{(1)},&
\map(J_{(2)})&=\sfrac{3}{2}\dual J_{(2)}-\sfrac{1}{2}J_{(2)},
\nln
\map (\dual J_{(0)})&=0,&
\map (\dual J_{(1)})&=\sfrac{3}{2}J_{(1)}+\sfrac{1}{2}\dual J_{(1)},&
\map (\dual J_{(2)})&=\sfrac{3}{2}J_{(2)}-\sfrac{1}{2}\dual J_{(2)}.
\end{align}
Again it generates $A(\lambda)$ in \eqref{eq:LaxZ3} via the exponential map \eqref{eq:expmap}.
This is most conveniently seen by splitting up 
the Maurer--Cartan form into chiral components
\[J =
  \half (J_{(2)}-\ast J_{(2)})
+ \half (J_{(1)}-\ast J_{(1)})
+ J_{(0)} 
+ \half (J_{(2)}+\ast J_{(2)}) 
+ \half (J_{(1)}+\ast J_{(1)}) .
\]
Note that $\map$ acts on these vectors by multiplication by 
$-2,-1,0,+1,+2$, respectively.
Consequently, one finds these numbers as exponents in \eqref{eq:LaxZ3}.

Finally, the operator that generates the 
Lax connection \eqref{eq:LaxZ4} 
for the $\Integer_4$-graded symmetric superspace model reads
\begin{align}
\map(J_{(0)})&=0,
&
\map(J_{(1)})&=-J_{(1)},
&
\map(J_{(2)})&=-2\, \dual J_{(2)} ,
\nln&
&
\map(J_{(3)})&=+J_{(3)},
&
\map(\dual J_{(2)})&=-2\, J_{(2)} .
\end{align}

\section{General Sigma Models}
\label{sec:derivation}

Above we have seen that for several classes of integrable
coset space sigma models we can construct the Lax connection by exponentiating 
a simple operator $\map$ acting globally on the Maurer--Cartan forms. 
It is conceivable that this procedure can be generalised 
to a wider class of models. 

Here we turn the logic of the previous section around, 
and start with an operator $\map$.
We will investigate which properties $\map$ has to satisfy 
such that the family of connections $A(\lambda)$ 
constructed via exponentiation of $\map$ \eqref{eq:expmap} becomes flat. 

The model will be a generic non-linear sigma model
in the framework presented above.
We will not make assumptions on the action nor the equations of motion.
In fact, these will follow from the flatness of $A(\lambda)$.
This will provide us with simple means to 
construct integrable sigma models and associated Lax connections
in terms of the operator $\map$.

\paragraph{Setup.}

In the above examples we have seen that 
the operator $\map$ commutes with dualisation $\dual$. 
In other words, it preserves chirality.
In the following we shall assume that this holds for more generic models.
Let us therefore define the chiral components $J^\pm$ 
of the Maurer--Cartan $J$ form as
\[
J^\pm:=\half (J\pm \dual J),
\qquad
J=J^+ + J^-.
\]
Likewise we split up the action of the operator $\map$ 
into two maps $\map^\pm$ acting only on a single copy of the Lie algebra $\alg{g}$, 
but not on the form algebra
\[
\map(J^\pm) = \map^\pm(J^\pm),
\qquad
\map^\pm : \alg{g} \to \alg{g}.
\]
The resulting candidate Lax connection takes the form
\[\label{eq:LaxChiral}
A(\lambda)=
e^{\lambda \map^+}(J^+) + e^{\lambda \map^-}(J^-),
\]
and the associated flatness condition reads
\[
e^{\lambda \map^+}(dJ^+)
+e^{\lambda \map^-}(dJ^-)
+\comm{e^{\lambda \map^+}(J^+)}{e^{\lambda \map^-}(J^-)}
=0.
\]
For convenience, we have defined the Lie bracket $\comm{\cdot}{\cdot}$ 
for $\alg{g}$-valued one-forms $X,Y$ as 
\[
\comm{X}{Y}:=X\wedge Y+Y\wedge X.
\]
%

\paragraph{Derivatives.}

We now expand the flatness condition to see what constraints 
must be imposed on the maps $\map^\pm$ and on the fields $J^\pm$.
From the leading order we merely obtain the Maurer--Cartan equation 
\[\label{eq:LaxMC}
d J^+ + d J^- +\comm{J^+}{J^-}=0.
\]
The first order yields an equation 
\[\label{eq:LaxEOM}
\map^+(dJ^+) + \map^-(dJ^-)
+\comm{\map^+(J^+)}{J^-}
+\comm{J^+}{\map^-(J^-)}=0
\]
which we can interpret as the equation of motion for the model.
Now the above two equations allow to solve
for the derivative terms
\[\label{eq:JpmSol}
dJ^\pm
=
\pm
(\map^+-\map^-)^{-1}
\bigbrk{
\map^\mp(\comm{J^+}{J^-})
-\comm{\map^+(J^+)}{J^-}
-\comm{J^+}{\map^-(J^-)}
}.
\]
Note that in principle this solution requires the map $\map^+-\map^-$ to be invertible.
This is in fact not the case for coset models.
For the time being this need not worry us, and we will discuss the implications 
further below. 

\paragraph{Integrability.}

We can now return to the flatness condition
and substitute the solution
\begin{align}
0=\mathord{}&
e^{\lambda \map^+}
(\map^+-\map^-)^{-1}
\map^-(\comm{J^+}{J^-})
-e^{\lambda \map^-}
(\map^+-\map^-)^{-1}
\map^+(\comm{J^+}{J^-})
\nln &
-(e^{\lambda \map^+}-e^{\lambda \map^-})
(\map^+-\map^-)^{-1}
\bigbrk{\comm{\map^+(J^+)}{J^-}+\comm{J^+}{\map^-(J^-)}}
\nln &
+\comm{e^{\lambda \map^+}(J^+)}{e^{\lambda \map^-}(J^-)}.
\end{align}
This expression is convenient for considering integrability 
because all the derivative terms have disappeared. 
Hence it is a plain algebraic equation 
for the maps $\map^\pm$ which 
must hold irrespectively of the dynamics of the model.
To improve the handling of the equation, let us introduce
a short hand notation. We observe that all maps 
eventually act on the term $\comm{J^+}{J^-}$ or its components. 
We introduce an index $0,1,2$ for $\map^\pm$ to specify which 
of the components it acts on: 
$1$ for $J^+$, $2$ for $J^-$
and $0$ for the overall Lie bracket $\comm{\cdot}{\cdot}$. 
The flatness condition reduces to
\begin{align}
0=\mathord{}&
\bigl[
e^{\lambda \map^+_0}(\map^+_0-\map^-_0)^{-1}\map^-_0
-e^{\lambda \map^-_0}(\map^+_0-\map^-_0)^{-1}\map^+_0
\nln &
-(e^{\lambda \map^+_0}-e^{\lambda \map^-_0})(\map^+_0-\map^-_0)^{-1}(\map^+_1+\map^-_2)
+e^{\lambda \map^+_1+\lambda \map^-_2}\bigr]\bigbrk{\comm{J^+}{J^-}}.
\end{align}
Quite curiously this equation has two solutions 
which hold for all values of $\lambda$:
namely $\map^\pm_0=\map^+_1+\map^-_2$,
i.e.\ when the operator $\map^\pm_0$ acts in precisely the same way as
the combination $\map^+_1+\map^-_2$ on $\comm{J^+}{J^-}$.
This is straightforwardly seen by substituting $\map_0^\pm$ for $\map^+_1+\map^-_2$.
It implies that the flatness equation can be written 
in either of the two forms
\[
\Xi^\pm(\lambda)(\map^\pm_0-\map^+_1-\map^-_2)
\bigbrk{\comm{J^+}{J^-}}=0
\]
for some operators $\Xi^\pm(\lambda)$.
Now we would like to combine the two expressions into something 
of the sort
\[
\Xi(\lambda)(\map^+_0-\map^+_1-\map^-_2)(\map^-_0-\map^+_1-\map^-_2)
\bigbrk{\comm{J^+}{J^-}}=0.
\]
This would require that the maps $\map^+$ and $\map^-$ commute. 
Commutativity indeed holds in the examples discussed above, 
but we can even deal with the more general case of 
$\comm{\map^+}{\map^-}\neq 0$. 
The flatness condition can then be written in two forms%
\footnote{This form can also be derived
by expanding the exponential form of the flatness condition
order by order in $\lambda$, but the derivation is rather tedious.}
\[
\Xi(\lambda)
\bigbrk{(\map^+_0-\map^-_0)\map^\mp_0(\map^+_0-\map^-_0)^{-1}-\map^+_1-\map^-_2}
\bigbrk{\map^\pm_0-\map^+_1-\map^-_2}\bigbrk{\comm{J^+}{J^-}}=0,
\]
The two forms are indeed equivalent as can be shown with some elementary 
algebraic manipulations. It means that the flatness condition can be
solved for all values of $\lambda$ 
when the following equation holds
\[
\bigbrk{(\map^+_0-\map^-_0)\map^\mp_0(\map^+_0-\map^-_0)^{-1}-\map^+_1-\map^-_2}
\bigbrk{\map^\pm_0-\map^+_1-\map^-_2}\bigbrk{\comm{J^+}{J^-}}=0.
\]
It is actually also a necessary condition for any $\lambda$-independent solution 
because the equation coincides with the $\order{\lambda^2}$ contribution 
to the flatness condition. In other words, a connection $A(\lambda)$ 
of the form $\eqref{eq:LaxChiral}$ 
that is flat up to terms of order $\order{\lambda^3}$ is actually exactly flat
for all $\lambda$.

\paragraph{Constraints.}

Now let us return to the case when the map $\map^+-\map^-$ is not invertible.
It implies that there exists some non-trivial map $\cmap:\alg{g}\to\alg{g}$ 
such that
\[\label{eq:kappacond}
\cmap(\map^+-\map^-)=0
\]
We can apply this to the equation of motion \eqref{eq:LaxEOM} 
where we make use of the Maurer--Cartan equations \eqref{eq:LaxMC}
\[\label{eq:constraint}
\cmap_0(\map^\pm_0-\map^+_1-\map^-_2)(\comm{J^+}{J^-})=0
\]
Here the map $\cmap$ has removed all derivative terms, 
and what remains is a constraint equation of motion 
for any choice of $\cmap$ satisfying \eqref{eq:kappacond}.
These constraints ensure that the inverse of 
$(\map^+-\map^-)$ is well-defined in the solution \eqref{eq:JpmSol} for $dJ^\pm$.
In the flatness condition we must allow for
appearance of the constraints 
\[\label{eq:MapCond}
\bigbrk{(\map^+_0-\map^-_0)\map^\mp_0(\map^+_0-\map^-_0)^{-1}-\map^+_1-\map^-_2+\cmap_0}
\bigbrk{\map^\pm_0-\map^+_1-\map^-_2}(\comm{\gen{T}_1}{\gen{T}_2})=0.
\]
For a suitable choice of $\cmap$ satisfying \eqref{eq:kappacond}
this equation must now hold irrespectively of the field configuration $J^\pm$.
We have therefore replaced $J^\pm$ by two generators $\gen{T}_1,\gen{T}_2$ 
of the Lie algebra $\alg{g}$.
Note that the role of $\cmap$ in \eqref{eq:MapCond} is twofold:
On the one hand, it parametrises the ambiguity in defining 
the inverse of $(\map^+-\map^-)$. On the other hand,
it removes terms which vanish by virtue of the constraints.
The purely algebraic equation for $\map^\pm$ 
guarantees flatness of the family of connections $A(\lambda)$
defined in \eqref{eq:LaxChiral}. 

Let us finally write out equation \eqref{eq:MapCond} more explicitly:
The Lax connection \eqref{eq:LaxChiral} is flat if 
for some choice of $\cmap$ satisfying \eqref{eq:kappacond}
the following condition holds
for all $\gen{T}_1,\gen{T}_2\in \alg{g}$ 
\begin{align}
0=\mathord{}&
\bigbrk{(\map^+-\map^-)\map^\mp(\map^+-\map^-)^{-1}+\cmap}\map^\pm\comm{\gen{T}_1}{\gen{T}_2}
\nln&
-\bigbrk{(\map^+-\map^-)\map^\mp(\map^+-\map^-)^{-1}+\cmap+\map^\pm}\bigbrk{\comm{\map^+\gen{T}_1}{\gen{T}_2}+\comm{\gen{T}_1}{\map^-\gen{T}_2}}
\nln&
 +\comm{\map^+\map^+\gen{T}_1}{\gen{T}_2}
+2\comm{\map^+\gen{T}_1}{\map^-\gen{T}_2}
 +\comm{\gen{T}_1}{\map^-\map^-\gen{T}_2}
.
\end{align}
%

\paragraph{Action.}

A map $\map^\pm$ satisfying \eqref{eq:MapCond} not only defines
a Lax connection, but also the Maurer--Cartan equations
\eqref{eq:LaxMC} and the equations of motion \eqref{eq:LaxEOM}.
While the Maurer--Cartan equations are basic ingredients of the
sigma model, not all conceivable equations of motion 
can be generated from an action.
For a sigma model action with Wess--Zumino term 
we can assume the general form 
\[
S=\sfrac{1}{2}\int \tr J \wedge \bar\map(J)
+\sfrac{1}{2}\int \tr J\wedge \bar\map_*(\dual J)
-\frac{\alpha}{3}\int_3 \tr J\wedge J\wedge J,
\]
where $\bar\map$ and $\bar\map_*$ are two linear maps 
on the Lie algebra.
Note that we can assume that 
$\bar\map$ is antisymmetric w.r.t.\
the Cartan--Killing form $\tr \gen{T}^a\gen{T}^b$,
while $\bar\map_*$ is symmetric
\[
\bar\map^\trans=-\bar\map,
\qquad
\bar\map_*^\trans=\bar\map_*.
\]
The resulting equations of motion read
\[
\bar\map(dJ)
+\bar\map_*(d\dual J)
+\comm{J}{\bar\map(J)}
+\comm{J}{\bar\map_*(\dual J)}
+\half\alpha \comm{J}{J}=0.
\]
Converted to the chiral components $J^\pm$ it takes precisely the form in \eqref{eq:LaxEOM}
with 
(note that the overall scale of the operator $\map^\pm$ does not matter)
\[\label{eq:MapAction}
\map^\pm=\bar\map\pm \bar\map_*+\alpha.
\]
In particular this implies that $\map^\pm$ are related by transposition
up to a constant defining the Wess--Zumino term
\[\label{eq:MapTrans}
(\map^+)^\trans = -\map^- + 2\alpha.
\]
Curiously the map $\map^\pm$ therefore defines immediately
the action of a sigma model.

\section{Applications}
\label{sec:application}

In the following we apply the above results to specific coset space sigma models.

\paragraph{$\Integer_N$-Coset Model.}

Integrable sigma models on $\Integer_N$-coset spaces were
investigated in \cite{Young:2005jv}.
They generalise the coset space models with $\Integer_2$ and $\Integer_3$ 
symmetry discussed above.
The general form of the action was found to be
\[
S=
\frac{1}{2}\sum_{k=1}^N \int \tr J_{(k)}\wedge \dual J_{(N-k)}
+\frac{1}{2}\sum_{k=1}^N \int \tr \lrbrk{1-\frac{2k}{N}} J_{(k)}\wedge J_{(N-k)}.
\]
It is convenient to cast the action into a chiral form
\[
S=-\sum_{k=1}^N \frac{2k}{N} \int \tr J_{(k)}^+\wedge J_{(N-k)}^-.
\]
The resulting equations of motion in a chiral form read
\[
dJ_{(0)}+\sum_{j=0}^{N-1} \comm{J_{(k)}^+}{J_{(N-k)}^-}=
dJ_{(k)}^+ + \sum_{j=k}^{N-1} \comm{J_{(j)}^+}{J_{(N-j+k)}^-}=
dJ_{(k)}^- + \sum_{j=1}^{k} \comm{J_{(k-j)}^+}{J_{(j)}^-}=0.
\]
By means of \eqref{eq:MapAction} we can read off the maps $\map^\pm$ 
from the action
\begin{align}\label{eq:sigmaZN}
\map^+(\gen{T}_{(k)})&=-k\, \gen{T}_{(k)},
&\mbox{for }&
k=0,\ldots,N-1,
\nln
\map^-(\gen{T}_{(k)})&=(N-k)\, \gen{T}_{(k)},
&\mbox{for }&
k=1,\ldots,N.
\end{align}
Note that maps $\map^\pm$ are related by minus transposition 
in agreement with to \eqref{eq:MapTrans}.
This translates to the relation $\map^+_{(N-k)}=-\map^-_{(k)}$
for the eigenvalues.
Let us now consider the integrability condition \eqref{eq:MapCond}.
First of all, $\map^+$ commutes with $\map^-$.
Furthermore the combination $\map^+-\map^-$ is invertible
except on $\alg{g}_{(0)}$.
This means there is exists a non-trivial $\cmap$
satisfying \eqref{eq:kappacond},
but conveniently it solves \eqref{eq:constraint}
independently of the field $J$;
hence there are no constraints. 
The integrability condition simplifies to
\[
\bigbrk{\map^+_0-\map^+_1-\map^-_2}
\bigbrk{\map^-_0-\map^+_1-\map^-_2}(\comm{\gen{T}_{(j)}}{\gen{T}_{(k)}})=0.
\]
Here we set $j=0,\ldots N-1$ and $k=1,\ldots,N$ such that we can substitute 
the map's eigenvalues
\[
\bigbrk{-(j+k-N\delta_{j+k\geq N})+j-(N-k)}
\bigbrk{N-(j+k-N\delta_{j+k> N})+j-(N-k)}=0.
\]
Note that for the eigenvalues of $\map^\pm_0$ we have shifted the
grading $j+k$ of $\comm{\gen{T}_{(j)}}{\gen{T}_{(k)}}$
by a suitable factor of $N$ to recover the
range as defined in \eqref{eq:sigmaZN}.
The integrability condition reduces to 
\[
-N^2\bigbrk{1-\delta_{j+k\geq N}} \delta_{j+k> N}=0,
\]
which is a true statement. 
Hence the maps $\map^\pm$ generate a Lax connection
as expected. Its form agrees with the results in \cite{Young:2005jv}
\[
A(\lambda)=
\sum_{k=0}^{N-1} e^{-k\lambda} J_{(k)}^+
+\sum_{k=0}^{N-1} e^{k\lambda} J_{(N-k)}^-.
\]
%

\paragraph{$\Integer_4$ Symmetric Superspace Model.}

Next let us consider symmetric superspace models
with $\Integer_4$ symmetry \cite{Bena:2003wd}.%
\footnote{Supersymmetry is in fact dispensable for our treatment.
We can safely treat the model as a bosonic $\Integer_4$-coset model
with constraints.}
These models are not in the above class of $\Integer_N$-coset spaces.
What makes them interesting is that some 
of their equations of motion are constraints.
When used in the context of supersymmetric string theory, 
these constraints are related to kappa symmetry, i.e.\
local supersymmetry on the worldsheet. 

Their action generalises the action of symmetric space sigma models 
as follows
\[\label{eq:superZ4action}
S=-\int \tr J_{(2)}^+\wedge J_{(2)}^- 
  -\half \int \tr J_{(1)}^+\wedge J_{(3)}^-
  +\half \int \tr J_{(3)}^+\wedge J_{(1)}^-
.
\]
The resulting Maurer--Cartan equations together with the 
equations of motion \eqref{eq:Z4eqs} read in chiral form
\begin{align}
\label{eq:superZ4EOM}
0&=d J_{(0)}
+\half\comm{J_{(0)}}{J_{(0)}} 
+\comm{J_{(1)}}{J_{(3)}} 
+\half\comm{J_{(2)}}{J_{(2)}} ,
\nln
0&=d J_{(1)}
+\comm{J_{(1)}}{J_{(0)}} 
+\comm{J_{(3)}}{J_{(2)}} ,
\nln
0&=d J_{(2)}^+ +\comm{J_{(2)}^+}{J_{(0)}} + \half\comm{J_{(1)}}{J_{(1)}} ,
\nln
0&=d J_{(2)}^- +\comm{J_{(2)}^-}{J_{(0)}}  + \half\comm{J_{(3)}}{J_{(3)}},
\nln
0&=d J_{(3)}
+\comm{J_{(3)}}{J_{(0)}} 
+\comm{J_{(1)}}{J_{(2)}}.
\end{align}
Note that there are no equations of motion for the odd
components of $J$. Their equations are replaced by the two constraints 
\[
\label{eq:superZ4Constr}
\comm{J_{(2)}^+}{J_{(1)}}=\comm{J_{(2)}^-}{J_{(3)}}=0.
\]
From the action \eqref{eq:superZ4action} or from the equations of motion \eqref{eq:superZ4EOM}
we can straight-forwardly read off 
the map $\map^\pm$
\[
\map^\pm(\gen{T}_{(0)})=0,
\qquad
\map^\pm(\gen{T}_{(1)})=-\gen{T}_{(1)},
\qquad
\map^\pm(\gen{T}_{(2)})=\mp 2 \gen{T}_{(2)} ,
\qquad
\map^\pm(\gen{T}_{(3)})=\gen{T}_{(3)}.
\]
By exponentiation one obtains the model's Lax connection 
\eqref{eq:LaxZ4}
\[
A(\lambda) = 
 e^{-2\lambda} J_{(2)}^+
+ e^{-\lambda} J_{(1)} 
+J_{(0)}
+ e^\lambda J_{(3)}
+ e^{2\lambda} J_{(2)}^-,
\]
and it is easy to verify 
flatness using the above equations of motion and constraints.

Let us nevertheless consider the integrability condition
\eqref{eq:MapCond} in detail, to see how the constraints work.
The matrices $\map^\pm$ act as factors 
on the generators $\gen{T}_{(k)}$
\[
\begin{array}{c|rrrr}
k   &  0 &  1 &  2 &  3 \\\hline
\map^+_{(k)} &\phantom{+} 0 & -1 & -2 & +1
\\
\map^-_{(k)} & 0 & -1 & +2 & +1
\end{array}
\]
First of all, we compute the eigenvalues of $\map^+_1+\map^-_2-\sigma^\pm_0$
on $\comm{\gen{T}_{(k)}}{\gen{T}_{(j)}}$. 
The result $\map^+_{(k)}+\map^-_{(j)}-\map^-_{(k+j)}$ 
is displayed in the following tables
%
%
%
\[\label{eq:superZ4matrix}
\begin{array}{c|rrrr}
j\backslash k  
   &  0 &  1 &  2 &  3 \\\hline
1  &  0 &  0 & -4 &  0 \\
2  & +4 &  0 &  0 & +4 \\
3  &  0 &  0 &  0 & +4 \\
0  &  0 & \phantom{+}0 &  0 &  0 \\
\end{array}
\qquad
\qquad
\begin{array}{c|rrrr}
j\backslash k  
   &  0 &  1 &  2 &  3 \\\hline
1  &  0 & -4 & -4 &  0 \\
2  &  0 &  0 &  0 & +4 \\
3  & \phantom{+}0 &  0 &  0 &  0 \\
0  &  0 &  0 & -4 &  0 \\
\end{array}
\]
We observe that many terms have cancelled right away.
Next we multiply the two tables, element by element.
Only two terms survive: $16$ at $(j,k)=(1,2)$ and $(j,k)=(2,3)$.
These correspond to the constraints \eqref{eq:superZ4Constr}.
More formally, we can multiply the above matrices 
\eqref{eq:superZ4matrix} by the
projectors $\cmap^1$ and $\cmap^3$
defined through their eigenvalues
\[
\begin{array}{c|rrrr}
k   &  0 &  1 &  2 &  3 \\\hline
\cmap^1_{(k)} & 0 & 1 & 0 & 0
\\
\cmap^3_{(k)} & 0 & 0 & 0 & 1
\end{array}
\]
both of which annihilate the combination $\map^+-\map^-$.
The resulting matrices have non-zero entries only at 
the positions of the constraints
$(j,k)=(1,2)$ and $(j,k)=(2,3)$,
such that the integrability constraint \eqref{eq:MapCond} is satisfied.

\paragraph{Principal Chiral Model.}

The principal chiral model is a sigma model on the Lie group manifold $\grp{G}$
\cite{Zakharov:1978pp}.
The action has two terms 
\[
S=\frac{\beta}{2}\int \tr J \wedge \dual J - \frac{\alpha}{3} \int_3 \tr J\wedge J \wedge J.
\]
The resulting equations of motion for $J$ read
\[
dJ+J\wedge J=0,\quad
\beta d\dual J +\alpha J\wedge J=0\qquad\mbox{or}\qquad
dJ^\pm  +\frac{\beta\pm\alpha}{2\beta}\, \comm{J^+}{J^-} =0.
\]
This model is integrable for all choices of parameters $\beta,\alpha$
and the following connection is flat for all $\lambda$
\[\label{eq:LaxPCM}
A(\lambda)
=\frac{1}{2\beta}\lrbrk{\beta-\alpha+(\beta+\alpha)e^{\lambda}}J^+
+\frac{1}{2\beta}\lrbrk{\beta+\alpha+(\beta-\alpha) e^{-\lambda}}J^-.
\]

Now we can try to apply our formula \eqref{eq:LaxChiral} 
to reproduce this connection.
According to \eqref{eq:MapAction} we should set $\map^\pm=\alpha\pm \beta$.
For generic $\alpha,\beta$ the condition \eqref{eq:MapCond} 
reduces to
\[
(\alpha+\beta)(\alpha-\beta)=0.
\]
The solution $\alpha=\pm\beta$ corresponds to the Wess--Zumino--Novikov--Witten model
\cite{Wess:1971yu,Novikov:1981,Witten:1983ar},%
\footnote{This result has been obtained earlier in 
discussions with T.~Bargheer and M.~Magro.}
and our formula \eqref{eq:LaxChiral},
$A(\lambda)=e^{\pm 2\beta\lambda}J^\pm+J^\mp$,
reproduces \eqref{eq:LaxPCM}, up to a rescaling of $\lambda$ by $2\beta$.

An alternative solution of our equation requires the presence of constraints. 
Here this is achieved by setting $\beta=0$,
where the equation of motion turns into the constraint
$dJ=J\wedge J=0$.
The condition \eqref{eq:MapCond} reduces to $(\alpha-\cmap)\alpha=0$
where according to \eqref{eq:kappacond} $\cmap$ is basically an 
unconstrained number, i.e.\ $\cmap=\alpha$. 
Our Lax connection \eqref{eq:LaxChiral} reads $A(\lambda)=e^{\alpha\lambda}J$ 
which is indeed flat subject to the constraint. 
The general solution \eqref{eq:LaxPCM} 
on the other hand is singular at $\beta=0$.
In any case, the constraint presumably makes this model rather uninteresting.

We have thus found an integrable model with a Lax connection 
which is not of the proposed form \eqref{eq:LaxChiral}. 
This is not surprising because \eqref{eq:LaxPCM} contains terms
proportional to $e^{\lambda}$, $e^{-\lambda}$ as well as $1=e^{0\lambda}$.
According to \eqref{eq:LaxChiral} there can be only two different 
exponents as $\map^\pm$ acts as a number in this case.

Nevertheless, it is surprising to see that the case $\alpha=0$ is not covered
by our formula. 
It is equivalent to a sigma model on the symmetric space
$\grp{G}=(\grp{G}\times\grp{G})/\grp{G}$
for which our formulation certainly applies using the map \eqref{eq:MapZ2}.
Here the deficit is due to the gauge fixing that eliminates
one of the two numerator groups. This gauge fixing does not commutes 
with our map $\map$.

To undo the gauge fixing, set 
\[
g=g_1 g_2^{-1},\qquad
J=g_2(J_1-J_2) g_2^{-1}.
\]
We can substitute this into the above action
to obtain 
\[
S=
  \frac{\beta}{2}\int \tr (J_1-J_2) \wedge \dual (J_1-J_2) 
- \frac{\alpha}{3} \int \tr J_1 \wedge J_2
- \frac{\alpha}{3} \int_3 \tr J_1\wedge J_1 \wedge J_1
+ \frac{\alpha}{3} \int_3 \tr J_2\wedge J_2 \wedge J_2
\]
Also the equations of motion follow by substitution
\[
\half\beta d\dual (J_1-J_2) 
+\half \beta \comm{J_2}{\dual (J_1-J_2)}
+\quarter \alpha \comm{J_1-J_2}{J_1-J_2}=0,
\]
and the Maurer--Cartan equations
read by construction
$dJ_1+J_1\wedge J_1=dJ_2+J_2\wedge J_2=0$.
To bring these into the framework of symmetric spaces,
use the Lie algebra $\alg{g}\oplus\alg{g}$ 
and set $J=(J_1,J_2)$ where the $\Integer_2$-automorphism 
exchanges the two Lie algebras.
In fact this case of a symmetric space is special because 
the the direct sum of two gauge algebras 
allows for two individual Wess--Zumino
coefficients $\alpha_1,\alpha_2$.
If chosen with opposite prefactors $\alpha_1=-\alpha_2=\alpha$,
invariance under the denominator group can be recovered.

A ($\alg{g}\oplus\alg{g}$)-valued Lax connection can now be written in the form
\[
A(\lambda)=J + (e^{\lambda}-1) \map^+ J^+ - (e^{-\lambda}-1) \map^- J^-,
\]
where $\map^\pm$ are operators acting on $\alg{g}\oplus\alg{g}$ 
as the $2\times 2$ matrix
\[
\map^\pm=\frac{1}{2\beta}\matr{cc}{\alpha\pm\beta&-\alpha\mp\beta\\\alpha\mp\beta&-\alpha\pm\beta}.
\]
The matrices are projectors, 
$(\map^\pm)^2=\pm\map^\pm$, 
and therefore they exponentiate according to the rule
\[
\exp(\lambda\map^\pm)=
1\pm(e^{\pm\lambda}-1)\map^\pm.
\]
Hence, the ($\alg{g}\oplus\alg{g}$)-valued Lax connection for the principal chiral model 
is indeed given by our formula \eqref{eq:LaxChiral}.
The algebraic form of the flatness condition \eqref{eq:MapCond} 
is satisfied (for any choice of $\cmap$).
Furthermore, the form of $\map^\pm$ agrees with \eqref{eq:MapAction} 
obtained from the action. 
In other words, our framework also fully applies to the principal chiral model
as long as a formulation with $(\alg{g}\oplus\alg{g})$-valued connections
is chosen.
We would like to point out that similar complications
with the original, gauge-fixed formulation
have been encountered in several other studies.

\paragraph{General $\Integer_2$-Models.}

Our framework allows to scan for integrable models. 
Let us consider a class of models 
where the Maurer--Cartan is split up 
according to a $\Integer_2$-automorphism of $\grp{G}$.
The most general action uses three constants $\alpha,\beta,\gamma$:
\[
S
=
\frac{\beta}{2}\int \tr J_{(0)}\wedge \dual J_{(0)}
+
\frac{\gamma}{2}\int \tr J_{(1)}\wedge \dual J_{(1)}
-
\frac{\alpha}{3}\int \tr J\wedge J\wedge J.
\]
The eigenvalues of the map $\map^\pm$ can be read off from \eqref{eq:MapAction}
\[
\map^+_{(0)}=+\beta+\alpha,
\qquad
\map^-_{(0)}=-\beta+\alpha,
\qquad
\map^+_{(1)}=+\gamma+\alpha,
\qquad
\map^-_{(1)}=-\gamma+\alpha.
\]
The corresponding equations of motion take the form
\begin{align}
0&=
dJ^+_{(0)}+dJ^-_{(0)}
+ \comm{J^+_{(0)}}{J^-_{(0)}}
+ \comm{J^+_{(1)}}{J^-_{(1)}},
\nln
0&=
dJ^+_{(1)}+dJ^-_{(1)}
+ \comm{J^+_{(0)}}{J^-_{(1)}}
+ \comm{J^+_{(1)}}{J^-_{(0)}},
\nln
0&=
\beta (dJ^+_{(0)}-dJ^-_{(0)})
+ \alpha\comm{J^+_{(0)}}{J^-_{(0)}}
+ \alpha\comm{J^+_{(1)}}{J^-_{(1)}},
\nln
0&=
\gamma(dJ^+_{(1)}-dJ^-_{(1)})
+ (\gamma-\beta+\alpha)\comm{J^+_{(1)}}{J^-_{(0)}}
+ (\beta-\gamma+\alpha)\comm{J^+_{(0)}}{J^-_{(1)}}.
\end{align}
Putting everything together, 
the integrability constraint \eqref{eq:MapCond} reads
\[
(\alpha-\beta)(\alpha+\beta)-\pi_0\alpha 
= (\alpha+2\gamma-\beta-\pi_1)(\alpha-\beta)
=(\alpha+\beta)(\alpha+\beta-2\gamma-\pi_1)=0
\]
Here the constants $\pi_0$ and $\pi_1$ are related to the 
appearance of constraints: 
The first can be nonzero, $\pi_0\neq 0$, only if $\beta=0$;
the second can be nonzero, $\pi_1\neq 0$, only if $\gamma=0$.
We should therefore distinguish the various cases. 

If we allow no constraints, the integrability constraint requires
$\beta=\gamma=\pm\alpha$. These are the parameters for the principal chiral model
which is discussed above.
The same applies to the case $\beta=\gamma=0$.
For $\beta=0$ we must set $\alpha=0$
which is the standard symmetric space model treated above.

For $\gamma=0$ we must set $\beta=\pm \alpha$.
In the case of $\beta=\alpha$,
the equations of motion read
\begin{align}
0&=
dJ_{(1)}
+ \comm{J^-_{(0)}}{J_{(1)}}
=\comm{J^+_{(0)}}{J_{(1)}},
\nln
0&=
dJ^+_{(0)}
+ \comm{J^+_{(0)}}{J^-_{(0)}}
+ \comm{J^+_{(1)}}{J^-_{(1)}}
=dJ^-_{(0)}.
\end{align}
The corresponding Lax connection is indeed flat
\[
A(\lambda)=e^{2\lambda}J^+_{(0)}+J^-_{(0)}+e^{\lambda}J_{(1)}.
\]
We have thus found one non-standard model in this class.

%
%
%

\paragraph{$D=2$, $\mathcal{N}=16$ Supergravity.}

Gravity theories dimensionally reduced to two dimensions
take a form similar to sigma models.
Moreover there are many interesting cases
where the model is integrable
due to the appearance of an infinite-dimensional symmetry algebra
as in \cite{Geroch:1972yt}. 
Let us discuss the case of maximal $\mathcal{N}=16$ supersymmetry 
in two dimensions.
It can be viewed as an $\grp{E}_8/\grp{SO}(16)$ 
coset model coupled to fermionic matter,
and a Lax connection was constructed in \cite{Nicolai:1987kz,Nicolai:1988jb}.
The coset model is based on a standard $\Integer_2$ symmetric space,
but it is interesting to see whether our construction of the Lax
connection can also be applied in the presence of matter.

We start with the Lax connection in chiral form given in \cite{Nicolai:1998gi}
\begin{align}
A^\pm (\lambda)=\mathord{}&
Q^\pm 
\pm \sfrac{i}{2} (e^{\mp 2\lambda}-1) (8\psi_2^\pm \cdot \psi^\pm \pm \chi^\pm \cdot \chi^\pm)
-2i(e^{\mp2\lambda}-1)^2 \psi_2^\pm\cdot\psi_2^\pm
\nln
&
+ e^{\mp \lambda} P^\pm
\mp i e^{\mp \lambda}(e^{\mp 2\lambda}-1) \psi_2^\pm\cdot\chi^\pm 
\end{align}
We have substituted the spectral parameter $\gamma=(e^{\lambda}-1)/(e^{\lambda}+1)$.
The fields $Q$ and $P$ are the grading $0$ and $1$ components of the 
standard Maurer--Cartan forms for the $\grp{E}_8/\grp{SO}(16)$ coset.
Furthermore, there are fermionic fields $\psi$, $\psi_2$ and $\chi$
which we do not need to define in detail here. 
Two such fields can be combined into an $\alg{e}_8$-valued form;
again we will not make this precise, but refer to the exact expression
given in \cite{Nicolai:1998gi}.

We observe that the Lax connection is again a Laurent polynomial in $e^\lambda$,
and it is conceivable that it is generated by our exponential construction \eqref{eq:expmap}.
This is indeed the case, if we define our map $\map$ to act as follows
\begin{align}
\map(Q^\pm) &= \mp i \chi^\pm\cdot \chi^\pm-8i \psi_2^\pm \cdot \psi^\pm,
\nln
\map(P^\pm) &= \mp P^\pm+2i \psi_2^\pm \cdot \chi^\pm,
\nln
\map(\psi^\pm) &= 2\psi_2^\pm,
\nln
\map(\chi^\pm) &= \mp \chi^\pm,
\nln
\map(\psi_2^\pm) &= \mp 2\psi_2^\pm.
\end{align}
Several comments are in order. 
Here the map does not simply act on the algebra $\alg{e}_8$,
which is quite obvious in the presence of matter.
It would be interesting to see how the action can now be formalised,
perhaps by enhancing the algebra.
Secondly, the fermions appear quadratically in the Lax connection itself. 
We should note that the map acts according to the Leibniz rule on products
of fields:
$\map(A\cdot B)=\map(A)\cdot B + A\cdot \map(B)$.
Finally, the spectral parameter $\lambda$ is position-dependant,
its derivatives are related to the dilaton field.
This feature appears to be no obstacle for our construction.

\section{Shift Operator}
\label{sec:shift}

At the end we can contemplate about the meaning of the operator $\map$. 
Consider $A(\lambda+\lambda')$, and split up the
exponent in \eqref{eq:expmap} in different ways to obtain
the relation
\[
A(\lambda+\lambda')=
\exp(\lambda'\map)\circ A(\lambda).
\]
The operator $\exp(\lambda'\map)$ thus shifts the parameter $\lambda$ 
by $\lambda'$. The operator $\map$ itself corresponds to an infinitesimal shift
\[
\frac{\partial}{\partial\lambda}\, A(\lambda)=
\map\circ A(\lambda).
\]
We can view this as a differential equation for the Lax connection.
Although the derivative of $A(\lambda)$ can certainly be expressed in some way, 
it is very remarkable that it is given by a $\lambda$-independent linear operator 
acting on $A(\lambda)$.

It is also useful to consider the following point of view: 
$A(\lambda)$ can be viewed as an element of the loop 
algebra $\alg{g}[e^{\lambda},e^{-\lambda}]$.
In this picture $\partial/\partial \lambda$ is the derivation 
of the loop algebra, 
and the operator $\map$ 
is equivalent to it. 

\section{Conclusions and Outlook}
\label{sec:concl}

The integrable structure of two-dimensional 
relativistic sigma models on coset spaces 
can be formulated in terms of a Lax connection $A(\lambda)$.
In this paper we have proposed a
construction for this one-parameter family 
of flat connections via exponentiating
a simple linear operator $\map$
acting on the Maurer--Cartan form $J$ of the model, $A(\lambda)=\exp(\lambda\map)J$. 
We have derived a quadratic constraint that this 
operator must satisfy such that the 
Lax connection is flat. 
Furthermore, the operator can be related directly
to the action and to the equations of motion. 
In that sense, our construction can be viewed
as an immediate integrability test for coset models. 
We can also turn the logic around, 
and use the construction to scan for integrable sigma models
by enumerating suitable operators $\map$. 
The action, equations of motion as well as the Lax connection 
all follow from the operator $\map$.

We have tested our construction for a large class
of models including 
symmetric (super)space models, $\Integer_N$-coset models,
principal chiral models as well as (super)gravity models.
It applies to all of these models,
but for the principal chiral model 
it is restricted to a coset space formulation.
However, there are also integrable sigma models 
where the construction does not seem to apply at all. 
For example, a sigma model on a squashed sphere 
was shown to be integrable \cite{Kawaguchi:2011pf}.
This model has two dual formulations of a Lax connection. 
One of them is equivalent to the principal chiral model, 
but in a gauge-fixed formulation to which our construction does not apply. 
The other one turns out to require at least 
four singular points in the spectral parameter plane. 
Conversely, our construction can only generate poles at 
two points, namely $z=e^\lambda=0,\infty$.
Hence our proposal does not apply to the sigma model on the squashed sphere.
This model can be viewed as a trigonometric integrable model
in analogy to the XXZ Heisenberg spin chain. 
The models which we have successfully tested are more reminiscent 
of rational integrable models analogous to the XXX chain. 
Therefore our construction most likely only applies to integrable
models of rational kind, but not of trigonometric kind. 
It would be interesting to see if and how the construction can be
generalised to trigonometric models.
Relativistic invariance would be another requirement that can be relaxed.
In this case, we expect that our construction can still be made to apply. 

Here we have only investigated the Lagrangian framework
of the sigma models. Concerning algebraic questions and quantisation it would
be desirable to establish our construction in the Hamiltonian framework. 
In fact, the operator $\map$ directly determines the Lagrangian, 
hence the Poisson structure follows from it. 
A logical next step would be to derive the algebra of monodromy matrices
which is usually governed by a pair of (r,s)-matrices
\cite{Maillet:1985fn}. 
Our expectation would be that they can also be formulated reasonably 
simply in terms of the operator $\map$. 

Ultimately we would like to understand the deeper meaning of
the operator $\map$. How is it related to the algebraic
formulation of integrability? 
The role as a shift operator for the 
spectral parameter $\lambda$ might be a first clue.

\paragraph{Acknowledgements.}

We are grateful to Till Bargheer and Marc Magro 
for collaboration at preliminary stages of this project. 

The work of NB is partially supported by grant no.\ 200021-137616 from
the Swiss National Science Foundation.

\begin{bibtex}
@article{Young:2005jv,
      author         = "Young, Charles A. S.",
      title          = "{Non-local charges, Z(m) gradings and coset space actions}",
      journal        = "Phys.Lett.",
      volume         = "B632",
      pages          = "559-565",
      doi            = "10.1016/j.physletb.2005.10.090",
      year           = "2006",
      eprint         = "hep-th/0503008",
      archivePrefix  = "arXiv",
      primaryClass   = "hep-th",
      SLACcitation   = "
}

@article{Bena:2003wd,
      author         = "Bena, Iosif and Polchinski, Joseph and Roiban, Radu",
      title          = "{Hidden symmetries of the AdS$_5$ $\times$ S$^5$ superstring}",
      journal        = "Phys.Rev.",
      volume         = "D69",
      pages          = "046002",
      doi            = "10.1103/PhysRevD.69.046002",
      year           = "2004",
      eprint         = "hep-th/0305116",
      archivePrefix  = "arXiv",
      primaryClass   = "hep-th",
      SLACcitation   = "
}

@article{Nicolai:1988jb,
      author         = "Nicolai, H. and Warner, N. P.",
      title          = "{The Structure Of $\mathcal{N}$ = 16 Supergravity In Two-Dimensions}",
      journal        = "Commun.Math.Phys.",
      volume         = "125",
      pages          = "369",
      doi            = "10.1007/BF01218408",
      year           = "1989",
      SLACcitation   = "
}

@article{Nicolai:1987kz,
      author         = "Nicolai, H.",
      title          = "{The Integrability Of $\mathcal{N}$ = 16 Supergravity}",
      journal        = "Phys.Lett.",
      volume         = "B194",
      pages          = "402",
      doi            = "10.1016/0370-2693(87)91072-0",
      year           = "1987",
      SLACcitation   = "
}

@article{Nicolai:1998gi,
      author         = "Nicolai, H. and Samtleben, H.",
      title          = "{Integrability and canonical structure of d = 2, $\mathcal{N}$ = 16
                        supergravity}",
      journal        = "Nucl.Phys.",
      volume         = "B533",
      pages          = "210-242",
      doi            = "10.1016/S0550-3213(98)00496-9",
      year           = "1998",
      eprint         = "hep-th/9804152",
      archivePrefix  = "arXiv",
      primaryClass   = "hep-th",
      reportNumber   = "AEI-061",
      SLACcitation   = "
}

@article{Eichenherr:1979ci,
      author         = "Eichenherr, H. and Forger, M.",
      title          = "{On the Dual Symmetry of the Nonlinear Sigma Models}",
      journal        = "Nucl.Phys.",
      volume         = "B155",
      pages          = "381",
      doi            = "10.1016/0550-3213(79)90276-1",
      year           = "1979",
      SLACcitation   = "
}

@article{Eichenherr:1979hz,
      author         = "Eichenherr, H. and Forger, M.",
      title          = "{More About Nonlinear Sigma Models On Symmetric Spaces}",
      journal        = "Nucl.Phys.",
      volume         = "B164",
      pages          = "528",
      doi            = "10.1016/0550-3213(80)90525-8,
                        10.1016/0550-3213(80)90525-8",
      year           = "1980",
      SLACcitation   = "
}

@article{Eichenherr:1981sk,
      author         = "Eichenherr, H. and Forger, M.",
      title          = "{Higher Local Conservation Laws For Nonlinear Sigma
                        Models On Symmetric Spaces}",
      journal        = "Commun.Math.Phys.",
      volume         = "82",
      pages          = "227",
      doi            = "10.1007/BF02099918",
      year           = "1981",
      SLACcitation   = "
}

@article{Pohlmeyer:1975nb,
      author         = "Pohlmeyer, K.",
      title          = "{Integrable Hamiltonian Systems and Interactions Through
                        Quadratic Constraints}",
      journal        = "Commun.Math.Phys.",
      volume         = "46",
      pages          = "207-221",
      doi            = "10.1007/BF01609119",
      year           = "1976",
      reportNumber   = "DESY 75/28",
      SLACcitation   = "
}

@article{Zakharov:1978pp,
      author         = "Zakharov, V. E. and Mikhailov, A. V.",
      title          = "{Relativistically Invariant Two-Dimensional Models in
                        Field Theory Integrable by the Inverse Problem Technique.}",
      journal        = "Sov.Phys.JETP",
      volume         = "47",
      pages          = "1017-1027",
      year           = "1978",
      SLACcitation   = "
}

@article{Luscher:1977rq,
      author         = {L\"uscher, M. and Pohlmeyer, K.},
      title          = "{Scattering of Massless Lumps and Nonlocal Charges in the
                        Two-Dimensional Classical Nonlinear Sigma Model}",
      journal        = "Nucl.Phys.",
      volume         = "B137",
      pages          = "46",
      doi            = "10.1016/0550-3213(78)90049-4",
      year           = "1978",
      reportNumber   = "DESY 77/65",
      SLACcitation   = "
}

@article{Brezin:1979am,
      author         = "Brezin, E. and Itzykson, C. and Zinn-Justin, Jean and
                        Zuber, J. B.",
      title          = "{Remarks About the Existence of Nonlocal Charges in
                        Two-Dimensional Models}",
      journal        = "Phys.Lett.",
      volume         = "B82",
      pages          = "442-444",
      doi            = "10.1016/0370-2693(79)90263-6",
      year           = "1979",
      SLACcitation   = "
}

@article{Schwarz:1995td,
      author         = "Schwarz, John H.",
      title          = "{Classical symmetries of some two-dimensional models}",
      journal        = "Nucl.Phys.",
      volume         = "B447",
      pages          = "137-182",
      doi            = "10.1016/0550-3213(95)00276-X",
      year           = "1995",
      eprint         = "hep-th/9503078",
      archivePrefix  = "arXiv",
      primaryClass   = "hep-th",
      reportNumber   = "CALT-68-1978",
      SLACcitation   = "
}

@article{Kawaguchi:2011pf,
      author         = "Kawaguchi, Io and Yoshida, Kentaroh",
      title          = "{Hybrid classical integrability in squashed sigma
                        models}",
      journal        = "Phys.Lett.",
      volume         = "B705",
      pages          = "251-254",
      year           = "2011",
      eprint         = "1107.3662",
      archivePrefix  = "arXiv",
      primaryClass   = "hep-th",
      reportNumber   = "KUNS-2350",
      SLACcitation   = "
}

@article{Eichenherr:1978qa,
      author         = "Eichenherr, H.",
      title          = "{SU(N) Invariant Nonlinear Sigma Models}",
      journal        = "Nucl.Phys.",
      volume         = "B146",
      pages          = "215-223",
      doi            = "10.1016/0550-3213(78)90439-X,
                        10.1016/0550-3213(78)90439-X",
      year           = "1978",
      SLACcitation   = "
}

@article{Zakharov:1979zz,
      author         = "Zakharov, V. E. and Shabat, A. B.",
      title          = "{Integration of nonlinear equations of mathematical
                        physics by the method of inverse scattering. II}",
      journal        = "Funct.Anal.Appl.",
      volume         = "13",
      pages          = "166-174",
      year           = "1979",
      doi            = "10.1007/BF01077483",
      SLACcitation   = "
}

@article{Zakharov:1974,
      author         = "Zakharov, V. E. and Shabat, A. B.",
      title          = "{A scheme for integrating the nonlinear equations of mathematical physics by the method of the inverse scattering problem. I}",
      journal        = "Funct.Anal.Appl.",
      volume         = "8",
      pages          = "226-235",
      year           = "1974",
      doi            = "10.1007/BF01075696",
      SLACcitation   = "
}

@article{Wess:1971yu,
      author         = "Wess, J. and Zumino, B.",
      title          = "{Consequences of anomalous Ward identities}",
      journal        = "Phys.Lett.",
      volume         = "B37",
      pages          = "95",
      year           = "1971",
      doi            = "10.1016/0370-2693(71)90582-X",
      SLACcitation   = "
}

@article{Witten:1983tw,
      author         = "Witten, Edward",
      title          = "{Global Aspects of Current Algebra}",
      journal        = "Nucl.Phys.",
      volume         = "B223",
      pages          = "422-432",
      doi            = "10.1016/0550-3213(83)90063-9",
      year           = "1983",
      reportNumber   = "PRINT-83-0262 (PRINCETON)",
      SLACcitation   = "
}

@article{Novikov:1981,
      author         = "Novikov, S. P.",
      title          = "{Multivalued functions and functionals. An analogue of the Morse theory}",
      journal        = "Sov. Math. Dokl.",
      volume         = "24",
      pages          = "222-226",
      year           = "1981",
}

@article{Witten:1983ar,
      author         = "Witten, Edward",
      title          = "{Nonabelian Bosonization in Two-Dimensions}",
      journal        = "Commun.Math.Phys.",
      volume         = "92",
      pages          = "455-472",
      doi            = "10.1007/BF01215276",
      year           = "1984",
      reportNumber   = "PRINT-83-0934 (PRINCETON)",
      SLACcitation   = "
}

@article{Geroch:1972yt,
      author         = "Geroch, Robert P.",
      title          = "{A Method for generating new solutions of Einstein's
                        equation. 2}",
      journal        = "J.Math.Phys.",
      volume         = "13",
      pages          = "394-404",
      doi            = "10.1063/1.1665990",
      year           = "1972",
      SLACcitation   = "
}

@article{Roiban:2000yy,
      author         = "Roiban, R. and Siegel, W.",
      title          = "{Superstrings on AdS$_5$ $\times$ S$^5$ supertwistor space}",
      journal        = "JHEP",
      volume         = "0011",
      pages          = "024",
      year           = "2000",
      eprint         = "hep-th/0010104",
      archivePrefix  = "arXiv",
      primaryClass   = "hep-th",
      reportNumber   = "YITP-SB-00-64",
      SLACcitation   = "
}

@article{Berkovits:1999zq,
      author         = "Berkovits, N. and Bershadsky, M. and Hauer, T. and
                        Zhukov, S. and Zwiebach, B.",
      title          = "{Superstring theory on AdS$_2$ $\times$ S$^2$ as a coset
                        supermanifold}",
      journal        = "Nucl.Phys.",
      volume         = "B567",
      pages          = "61-86",
      doi            = "10.1016/S0550-3213(99)00683-5",
      year           = "2000",
      eprint         = "hep-th/9907200",
      archivePrefix  = "arXiv",
      primaryClass   = "hep-th",
      reportNumber   = "IFT-P-060-99, HUTP-99-A044, MIT-CTP-2878",
      SLACcitation   = "
}

@article{Kallosh:1998zx,
      author         = "Kallosh, Renata and Rahmfeld, J. and Rajaraman, Arvind",
      title          = "{Near horizon superspace}",
      journal        = "JHEP",
      volume         = "9809",
      pages          = "002",
      year           = "1998",
      eprint         = "hep-th/9805217",
      archivePrefix  = "arXiv",
      primaryClass   = "hep-th",
      reportNumber   = "SLAC-PUB-7840, SU-ITP-98-36",
      SLACcitation   = "
}

@article{Metsaev:1998it,
      author         = "Metsaev, R. R. and Tseytlin, Arkady A.",
      title          = "{Type IIB superstring action in AdS$_5$ $\times$ S$^5$
                        background}",
      journal        = "Nucl.Phys.",
      volume         = "B533",
      pages          = "109-126",
      doi            = "10.1016/S0550-3213(98)00570-7",
      year           = "1998",
      eprint         = "hep-th/9805028",
      archivePrefix  = "arXiv",
      primaryClass   = "hep-th",
      reportNumber   = "FIAN-TD-98-21, IMPERIAL-TP-97-98-44, NSF-ITP-98-055",
      SLACcitation   = "
}

@article{Kallosh:1998nx,
      author         = "Kallosh, Renata and Rahmfeld, J.",
      title          = "{The GS string action on AdS$_5$ $\times$ S$^5$}",
      journal        = "Phys.Lett.",
      volume         = "B443",
      pages          = "143-146",
      doi            = "10.1016/S0370-2693(98)01281-7",
      year           = "1998",
      eprint         = "hep-th/9808038",
      archivePrefix  = "arXiv",
      primaryClass   = "hep-th",
      reportNumber   = "SU-ITP-98-47",
      SLACcitation   = "
}

@article{Beisert:2010jr,
      author         = "Beisert, Niklas and and others",
      title          = "{Review of AdS/CFT Integrability: An Overview}",
      journal        = "Lett.Math.Phys.",
      volume         = "99",
      pages          = "3-32",
      doi            = "10.1007/s11005-011-0529-2",
      year           = "2012",
      eprint         = "1012.3982",
      archivePrefix  = "arXiv",
      primaryClass   = "hep-th",
      reportNumber   = "AEI-2010-175, CERN-PH-TH-2010-306, HU-EP-10-87,
                        HU-MATH-2010-22, KCL-MTH-10-10, UMTG-270, UUITP-41-10",
      SLACcitation   = "
}

@article{Magro:2010jx,
      author         = "Magro, Marc",
      title          = "{Review of AdS/CFT Integrability, Chapter II.3: Sigma
                        Model, Gauge Fixing}",
      journal        = "Lett.Math.Phys.",
      volume         = "99",
      pages          = "149-167",
      doi            = "10.1007/s11005-011-0481-1",
      year           = "2012",
      eprint         = "1012.3988",
      archivePrefix  = "arXiv",
      primaryClass   = "hep-th",
      reportNumber   = "AEI-2010-132",
      SLACcitation   = "
}

@article{Maillet:1985fn,
      author         = "Maillet, Jean Michel",
      title          = "{Kac-Moody Algebra and Extended Yang-Baxter Relations in
                        the O(N) Nonlinear Sigma Model}",
      journal        = "Phys.Lett.",
      volume         = "B162",
      pages          = "137",
      doi            = "10.1016/0370-2693(85)91075-5",
      year           = "1985",
      reportNumber   = "PAR LPTHE 85-22",
      SLACcitation   = "
}

@article{Lax:1968fm,
      author         = "Lax, P. D.",
      title          = "{Integrals of Nonlinear Equations of Evolution and
                        Solitary Waves}",
      journal        = "Commun.Pure Appl.Math.",
      volume         = "21",
      pages          = "467-490",
      year           = "1968",
      doi            = "10.1002/cpa.3160210503",
      SLACcitation   = "
}

\end{bibtex}

\bibliographystyle{nb}
\bibliography{\jobname}

\end{document}